\documentclass[%
 reprint,
 amsmath,amssymb,
 aps,
]{revtex4-2}

\usepackage{graphicx}% Include figure files
\usepackage{dcolumn}% Align table columns on decimal point
\usepackage{bm}% bold math
\usepackage[utf8]{inputenc} % allow utf-8 input
\usepackage[T1]{fontenc}    % use 8-bit T1 fonts
\usepackage{hyperref}       % hyperlinks
\usepackage{url}            % simple URL typesetting
\usepackage{booktabs}       % professional-quality tables
\usepackage{amsfonts}       % blackboard math symbols
\usepackage{nicefrac}       % compact symbols for 1/2, etc.
\usepackage{microtype}      % microtypography
\usepackage{lipsum}
\usepackage{soul}
\usepackage{color}
\usepackage{float}
\usepackage{graphicx}
\usepackage{caption}
\usepackage{subcaption}
\usepackage{titlesec}
\usepackage{multirow}
\usepackage{amsmath}
\usepackage{soul,color}
\usepackage[section]{placeins}
\usepackage{array, makecell} %
\usepackage{float}
\usepackage[section]{placeins}
\usepackage{wrapfig}
\usepackage{enumitem}% http://ctan.org/pkg/enumitem
\usepackage[nodisplayskipstretch]{setspace}
\usepackage{comment}

\begin{document}

\preprint{APS/123-QED}

\title{Crystal structure prediction using age-fitness multi-objective genetic algorithm and coordination number constraints}% Force line breaks with \\
\thanks{Corresponding Author: Jianjun Hu.}%

\author{Wenhui Yang}
\affiliation{%
 School of Mechanical Engineering\\
  Guizhou University \\
  Guiyang China 550025 \\
%  This line break forced with \textbackslash\textbackslash
}

%  \altaffiliation[Also at ]{Physics Department, XYZ University.}%Lines break automatically or can be forced with \\
\author{Edirisuriya M. Dilanga Siriwardane}%
\affiliation{%
Department of Computer Science and Engineering\\
  University of South Carolina\\
  Columbia, SC 29201 \\
%  This line break forced with \textbackslash\textbackslash
}%

\author{Jianjun Hu}
 \email{jianjunh@cse.sc.edu}

 \homepage{http://www.cse.sc.edu/~jianjunh}
\affiliation{
 Department of Computer Science and Engineering\\
  University of South Carolina\\
  Columbia, SC, 29201, USA \\
%  This line break forced% with \\
}%
% \affiliation{
%  Third institution, the second for Charlie Author
% \affiliation{%
%  Authors' institution and/or address\\
%  This line break forced with \textbackslash\textbackslash
% }%

% \collaboration{CLEO Collaboration}%\noaffiliation

\date{\today}% It is always \today, today,

\begin{abstract}
Crystal structure prediction (CSP) has emerged as one of the most important approaches for discovering new materials. CSP algorithms based on evolutionary algorithms and particle swarm optimization have discovered a great number of new materials. However, these algorithms based on ab initio calculation of free energy are inefficient. Moreover, they have severe limitations in terms of scalability. We recently proposed a promising crystal structure prediction method based on atomic contact maps, using global optimization algorithms to search for the Wyckoff positions by maximizing the match between the contact map of the predicted structure and the contact map of the true crystal structure. However, our previous contact map based CSP algorithms have two major limitations: (1) the loss of search capability due to getting trapped in local optima; (2) it only uses the connection of atoms in the unit cell to predict the crystal structure, ignoring the chemical environment outside the unit cell, which may lead to unreasonable coordination environments. Herein we propose a novel multi-objective genetic algorithms for contact map-based crystal structure prediction by optimizing three objectives, including contact map match accuracy, the individual age, and the coordination number match. Furthermore, we assign the age values to all the individuals of the GA and try to minimize the age aiming to avoid the premature convergence problem. Our experimental results show that compared to our previous CMCrystal algorithm, our multi-objective crystal structure prediction algorithm (CMCrystalMOO) can reconstruct the crystal structure with higher quality and alleviate the problem of premature convergence.
\end{abstract}


\maketitle

% \tableofcontents

\section{Introduction}
The discovery and development of new materials are fundamental to the progress of technology. There are several promising approaches for exploring new materials including crystal structure predictions \cite{glass2006uspex,kvashnin2019computational,ryan2018crystal,hu2021contact}, generative machine learning models \cite{kim2020inverse,dan2020generative,bradshaw2019model,ren2020inverse}, inverse materials design \cite{zunger2018inverse,kim2020inverse}, and first-principles \cite{curtis2018gator,woodley2008crystal,maddox1988crystals} calculation based structural tinkering. Materials Genome Initiative attempts to use data-driven methods \cite{gomez2018automatic,jablonka2020big,hoffmann2019data} to help discover new material science research paradigms and accelerate the design and exploration of new materials. Since the structure of a material determines its many physical and chemical properties, crystal structure prediction is thus an important process for finding new materials. Global optimization and data mining are currently the two main crystal structure prediction methods \cite{2019Topology}. Data mining methods usually have higher requirements for crystal structure data and faster speed. However, because some data may not be complete and effective, it is easy to make mistakes. Evolutionary algorithms instead have become an important method for predicting crystal structures due to their excellent global optimization performance \cite{glass2006uspex,avery2019xtalopt}. The crystal structure prediction algorithms based on an evolutionary algorithm and particle swarm optimization have discovered many new materials \cite{oganov2011evolutionary,wang2020calypso}. However, these global optimization algorithms are usually based on ab initio calculations of free energies, which rely on expensive DFT calculations. This makes them inefficient and has many limitations in terms of scalability \cite{zhang2017materials,oganov2019structure}.

Compared to the vast chemical space of crystal materials, the known crystal structures (~200,000) as deposited in the ICSD and Materials Project database are quite limited. Recently, we proposed a generative machine learning model CubicGAN  \cite{zhao2021high} for automated generation of cubic crystal structures, allowing us to discover hundreds of new prototype cubic materials. However, that approach is currently only limited to generate cubic structures with special coordinates. Another strategy to generate hypothetical crystal structures is to first generate the compositions, using our composition generative machine learning model (MatGAN) \cite{dan2020generative}, which can generate hypothetical crystal material compositions by learning implicit chemical composition rules. Our model can be used to generate millions of new hypothetical materials compositions, with the potential to significantly expand the chemical design space for inorganic material design and large-scale computational screening \cite{kvashnin2019computational,he2020computational}. It is of great significance to analyze their physical and chemical properties, which depend on the availability of their structures. However, predicting the crystal structure of the given chemical composition is a major difficulty that has merits decades of research  \cite{oganov2019structure}. We previously proposed a contact map-based crystal structure prediction method AlphaCrystal \cite{hu2021alphacrystal}, which predicts the contact map \cite{0Contact}, space group, and lattice constants of the material through deep learning methods, and then uses global optimization algorithms such as GA and PSO to maximize the match between the contact map of the predicted structure and the predicted contact map by searching for the Wyckoff positions. Our experiments proved that the geometric constraints \cite{wei2019protein,zhu2017efficient,kuhlman2019advances} in the crystal structure help the crystal structure reconstruction. Compared with the crystal structure prediction method based on global free energy optimization \cite{oganov2011evolutionary,lyakhov2013new}, our method uses a large number of hidden geometric constraints, composition, and atomic configuration rules in known crystal structures, thus improving the sampling efficiency, which makes it suitable for large-scale crystal structure prediction. However, our previous crystal structure prediction method based on contact maps still has two major limitations in terms of its search capability. First, it only used the connections of atoms in the unit cell to predict the crystal structure, ignoring the chemical environment outside the unit cell, which may form an unreasonable coordination environment. As a result, some predicted crystal structures are very different from the target structures. Here we take the approach of physics informed machine learning, which incorporates physical principles into machine learning (ML) models \cite{kim2020inverse,dan2020generative,bradshaw2019model,noh2019inverse,ren2020inverse, dan2020generative}. We find that the polyhedral formed by cations and nearby anions can serve as an important optimization target to the machine learning framework of crystal materials \cite{2021SciA....7.1754B}. Therefore, we added the coordination number of the cation as an additional optimization objective to the previous contact map-based CSP algorithm, the CMCrystal.

In addition, during our usage of the crystal structure prediction method based on evolutionary algorithms such as CALYPSO  \cite{wang2015materials}, and CrySPY \cite{yamashita2018crystal}, we find both algorithms are easy to converge prematurely and fall into local optima. However, studies have shown that dividing the evolving population into different age groups can significantly improve the ability to obtain globally optimal solutions and avoid falling into local optimal solutions  \cite{hu2005hierarchical,hornby2006alps}. Therefore, in order to improve the performance of the evolutionary algorithm to reconstruct the crystal structure, we take ages of the individuals as an explicit optimization target in addition to the contact map match and the coordination number accuracy, leading to a multi-objective GA for CSP, the CMCrystalMOO algorithm  \cite{schmidt2011age}. This method evolves a population of non-dominated candidate solutions in the Pareto front considering the time (age) of the individuals in the population and their performance (fitness regarding contact map and coordination number prediction accuracy).
In the end, we constituted a multi-objective genetic algorithm with contact map, coordination number of the cation and ages of the individuals for crystal structure prediction. 

Our contributions can be summarized as follows:

\begin{itemize}

    \item 
    Compared to our previous contact map-based crystal structure prediction algorithm CMCrystal, we propose an additional optimization target, the coordination number of the cation, to comprehensively consider the chemical environment inside and outside the unit cell. The connection of atoms in the unit cell and the coordination number of the cation outside the unit cell need to be optimized. The reconstructed crystal structures tend to resemble the real crystal structure with this additional physics-informed optimization target. 
    \item 
    To address the common premature convergence issue of evolutionary algorithms in challenging optimization tasks, we introduced the ages of the individuals as one additional optimization objective in our optimization algorithm, which naturally leads to a multiobjective GA that takes the contact map accuracy, and the coordination number accuracy of the cation, and the age as the optimization objectives, which alleviates the problem of the genetic algorithm falling into local optima, and improves the global search ability of the genetic algorithm. 
    \item We evaluated our multi-objective genetic algorithm CMCrystalMOO on a set of crystal structures with extensive experiments to prove the effectiveness of our algorithm for reconstructing the structures from the contact map and coordination number of the cation. 
\end{itemize}

\section{Methods}

\subsection{Coordination number as optimization target for atomic coordinate reconstruction}

In a crystal structure,  the coordination number of a central atom or ion is the number of atoms or ions directly adjacent to it. It forms a coordination polyhedron when the neighboring atoms/ions are connected to a central atom/ion. In many cases, the radius ratio of the atom pair determines the coordination number of the cation in ionic crystals. The contact map records the connection relationship between all atoms in the unit cell and captures the interaction between atoms.

\begin{figure*}[htb] 
    \begin{subfigure}[ht]{0.3\textwidth}
        \includegraphics[width=\textwidth]{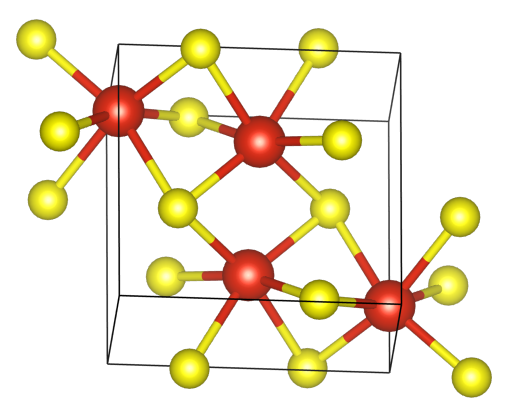}
        \caption{Structural display with complete coordination info for all atoms within unit cell (20 atoms, 24 bonds, 4 polyhedra)}
        \vspace{-3pt}
        \label{fig:V4S4_true}
    \end{subfigure}\hfill
    \begin{subfigure}[t]{0.23\textwidth}
        \includegraphics[width=\textwidth]{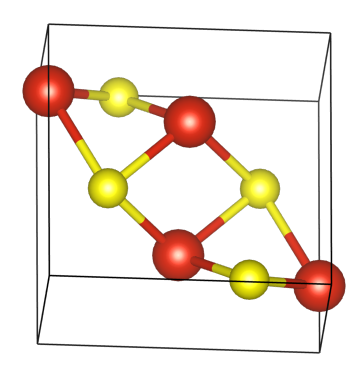}
        \caption{The connections of atoms within the unit cell(8 atoms, 10 bonds, 0 polyhedra)}
        \vspace{-3pt}
        \label{fig:V4S4_cell}
    \end{subfigure}\hfill
    \begin{subfigure}[t]{0.3\textwidth}
        \includegraphics[width=\textwidth]{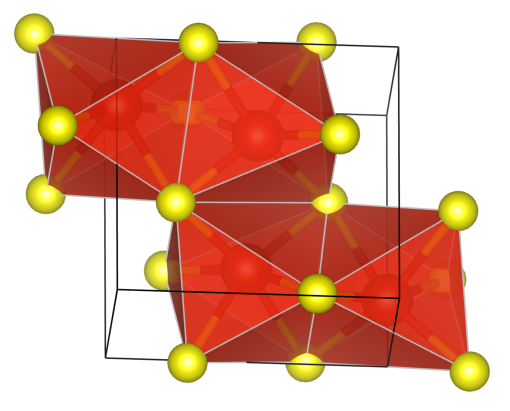}
        \caption{polyhedron representation }
        \vspace{-3pt}
        \label{fig:V4S4_poly}
    \end{subfigure}
    \caption{Different views of the crystal structure of  V\textsubscript{4}S\textsubscript{4}}
    \label{fig:contact}
\end{figure*}

Figure \ref{fig:contact} shows the crystal structure of  V\textsubscript{4}S\textsubscript{4}  with 20 atoms, 24 bonds, 4 polyhedra. There are 8 of 20 atoms in the unit cell, while the remaining atoms are the equivalent atoms of the S atoms in the neighbor unit cells due to the structure periodicity. Our previous contact map-based crystal structure prediction method only considers the connections of the 8 atoms within the unit cell, ignoring the coordination environment outside the unit cell, as shown in Figure\ref{fig:V4S4_cell}. So, we added the optimization goal of the coordination number of the cation to the previous method of crystal structure prediction based on contact map.

%\subsection{The contact map based crystal structure prediction problem}

\subsection{Age-fitness based multi-objective genetic algorithm for crystal structure prediction}

In our previous work \cite{hu2021contact}, the crystal structure prediction problem can be mapped to two related problems: 1) prediction of the contact map of atoms; 2) the atomic coordinate reconstruction from the contact map using global optimization algorithms. We have applied both genetic algorithms and differential evolution algorithms \cite{li2021composition} for the coordinate reconstruction. However, we find that in both algorithms, the evolving population can easily get trapped in local optima due to the premature convergence, when the diversity of the population decreases dramatically after a few generations, which leads to the loss of search capability. It is thus important to introduce prevention techniques to avoid or ameliorate the premature convergence issue.

There are several techniques that address the premature convergence issue, including the well known fitness sharing or niching techniques \cite{sareni1998fitness}, which aims to maintain population diversity. Additional research \cite{hu2005hierarchical} shows that a pipeline of new genetic materials is needed to maintain sustainable evolutionary search. Based on this idea, hierarchical fair competition algorithms \cite{hu2005hierarchical} , age-layered population search algorithm, multi-objective evolutionary algorithms \cite{schmidt2011age} have been proposed which have demonstrate much stronger evolutionary search capability for challenging global optimization problems. Here we choose the age-fitness multi-objective EA framework (AFMOEA) for building a more powerful genetic algorithm for atomic coordinate reconstruction from contact map and the coordination number of the cation. By optimizing the fitness function and minimizing the ages of the individuals, the age-fitness multi-objective algorithm implicitly maintains the population diversity and a genetic pipeline for sustainable evolutionary search.

\begin{figure}[ht]
  \centering
  \includegraphics[width=0.9\linewidth]{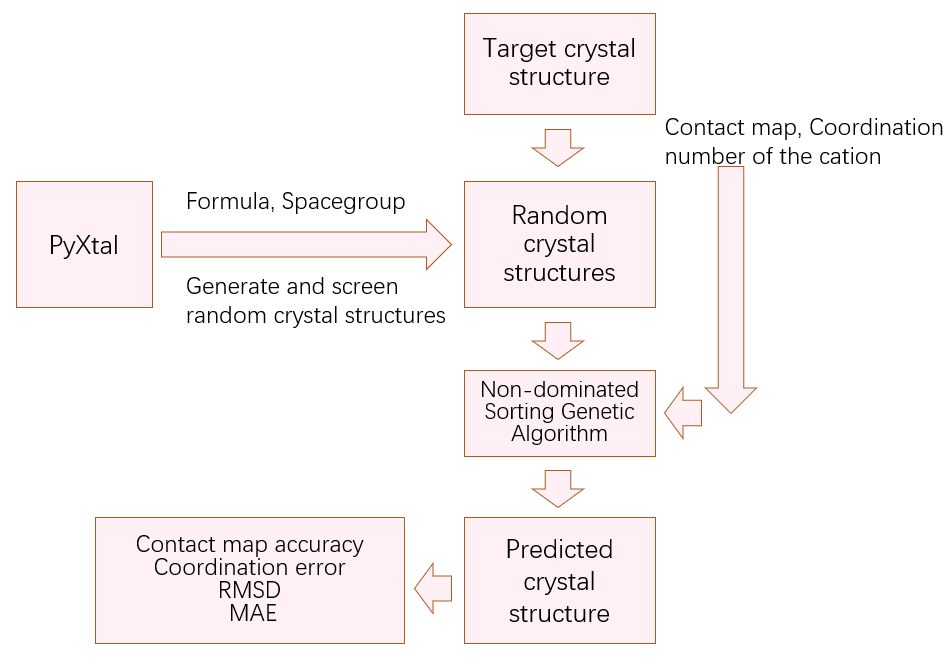}
  \caption{Prediction framework of multi-objective genetic algorithm.}
  \label{fig:framework}
\end{figure}

Here we propose a multi-objective crystal structure prediction algorithm based on contact map, coordination number of the cation and ages of individuals as shown in Figure \ref{fig:framework}. First, we use the PyXtal library \cite{fredericks2021pyxtal} to generate 50 random crystal structures based on the given formula and the space group. We then select five suitable random crystal structures, and use the multi-objective crystal structure prediction algorithm to search the Wyckoff positions coordinates by approximating the target contact map, the coordination number of the cation and minimizing the ages of individuals. The goal is to make the contact map of the optimized structure and the coordination number of the cation match as much as possible with the contact map of the real structure and the coordination number of the cation. We use the contact map accuracy, the coordination number error of the cation, the root mean square distance (RMSD) and the mean absolute error (MAE) between the predicted Wyckoff positions of the crystal structure and the Wyckoff positions of the target structure to evaluate the reconstructed crystal structure. 
 
\subsection{3D crystal structure reconstruction algorithm}

We use NSGA-II: Non-dominated Sorting Genetic Algorithm in the pymoo framework \cite{pymoo} to search for the Wyckoff positions guided by the contact map, the coordination number of the cations and the ages of the individuals to realize the prediction of the crystal structure. Compared with traditional genetic algorithms, NSGA-II has improved mating and survival selection. In NSGA-II, first, individuals are selected frontwise. Because not all individuals are allowed to survive, individuals on the front lines need to be divided and only representative individuals will be selected for next generation. In this split front, the candidate solutions are selected based on the crowding distance (the Manhattan Distance in the target space). In addition, NSGA-II uses binary tournament mating selection. Each individual first compares the rank and then the crowded distance.

In order to alleviate the premature convergence issue of NSGA-II, we added an additional optimization objective, the ages of individuals. The specific process is as follows: the age of an individual is based on generations. All randomly initialized individuals start with a age of one . Since then, the age of the individual will increase by one for each surviving generation. In the process of crossover and mutation, age is inherited as the maximum age of parents. In addition, we added 5\% of random individuals to the population of next generation to ensure the diversity of the population. The population of candidate solutions can be mapped to the two-dimensional plane of age and fitness. The goal of the multi-objective optimization is to find the non-dominated Pareto frontier of a given problem domain. Here, our target is to identify candidate structures with the smallest age and the greatest fitness, which leads to selection pressure to let old-mediocre candidates become extinct during the evolutionary process. 

In order to verify the performance whether adding the ages of individuals as additional optimization target to the NSGA-II that takes the contact map match and the coordination number match of the cations as the optimization targets, we analyze the process of searching the complex Li\textsubscript{4}Fe\textsubscript{4}F\textsubscript{16} Wyckoff positions with and without the age objective. We use the hypervolume to compare the search capability, which is a well-known performance indicator for multi-objective optimization  \cite{guerreiro2020hypervolume}. Calculation of hypervolume requires defining a reference point that shall be larger than the maximum value of the Pareto front. We select the reference point(0,16) with the contact map accuracy to be 0, and the max coordination number error to be 16. 

\begin{figure}[ht]
  \centering
  \includegraphics[width=0.8\linewidth]{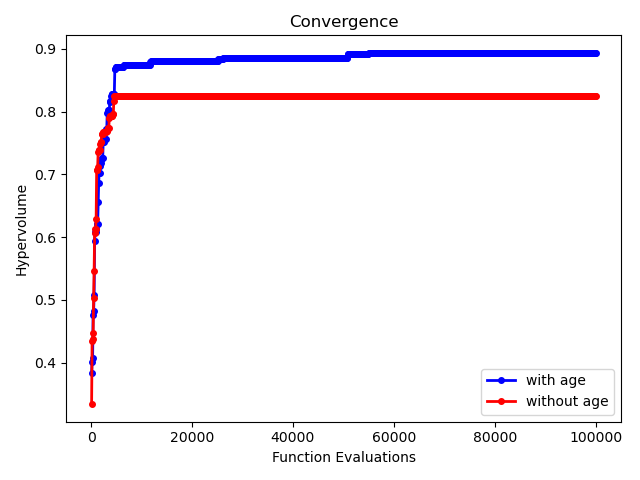}
  \caption{Hypervolume comparison of  multi-objective genetic algorithms with and without age. The age-enhanced GA has shown higher search performance.}
  \label{fig:convergence}
\end{figure}

As shown in Figure \ref{fig:convergence}, at the beginning, the performance of the multi-objective genetic algorithm with age and without age is close. With the increase of generation, the performance of multi-objective genetic algorithm with age is higher than the one without age, and there will be a gradual improvement in performance later. Experimental results show that our multi-objective genetic algorithm with age can effectively alleviate the premature convergence problem and improve the algorithm's global search ability. 
For all the NSGA-II based experiments, we set the population size to 100,the number of generations to 1000 and the crossover probability to 0.9.
\subsection{Evaluation metrics}

If the distance between the two atoms in the unit cell is within the range of $[Min.length,Max.length]$, it means there is a bond between them and the corresponding contact map position is set to 1, else it is set to 0. The contact map fitness function of the NSGA-II is as follows: 

\begin{equation}
\operatorname{fitness}_{opt}=\frac{2|A \cap B|}{|A|+|B|} \approx\frac{2 \times A \bullet B}{\operatorname{Sum}(A)+\operatorname{Sum}(B)}
\end{equation}

where $A$ is the predicted contact map and $B$ is the target contact map, both only contain 1/0 entries.  $A \cap B$  represents the common elements between A and B, |g| denotes the number of elements in a contact map, Sum(g) denotes the sum of all contact map elements. 

In our problem, the coordination number of the cation is the coordination number in the unit cell and outside the unit cell of the cation. Since the contact map has recorded the atomic connection in the unit cell, it is equivalent to include the coordination number in the unit cell of the cation. Therefore, we focus on analyzing the coordination number outside the unit cell of the cation. The coordination number error of the cation (to be precise, the coordination number outside the unit cell error of the cation) is defined as follows,

\begin{equation}
\operatorname{coordi-number-error-of- cation}={\operatorname{Sum}(|A-B|)}
\end{equation}

Where A is the predicted coordination number outside the unit cell of all cations, and B is the actual coordination number outside the unit cell of all cations. The value of A can be calculated as follows: in the process of searching Wyckoff positions in the optimization algorithm, first find out the equivalent positions outside the unit cell of the anion through symmetric operations, and then determine the cation and the anion outside the unit cell according to the atomic distance threshold in the Vesta software whether to pair or not.

To evaluate the reconstruction performance of NSGA-II, we can use the contact map accuracy and coordination number error of the cation as evaluation criteria. Moreover, we define the root mean square distance (RMSD) and mean absolute error (MAE) of two structures as below:
\begin{small}
\begin{equation}
    \begin{aligned}
\mathrm{RMSD} &=\sqrt{\frac{1}{n} \sum_{i=1}^{n}\left\|v_{i}-w_{i}\right\|^{2}} \\
&=\sqrt{\frac{1}{n} \sum_{i=1}^{n}\left(\left(v_{i x}-w_{i x}\right)^{2}+\left(v_{i y}-w_{i y}\right)^{2}+\left(v_{i z}-w_{i z}\right)^{2}\right)}
\end{aligned}
\end{equation}
\end{small}

\begin{small}
\begin{equation}
        \begin{aligned}
\mathrm{MAE} &=\frac{1}{n} \sum_{i=1}^{n}\left\|v_{i}-w_{i}\right\| \\
&=\frac{1}{n} \sum_{i=1}^{n}\left(\|v_{i x}-w_{i x}\|+\|v_{i y}-w_{i y}\|+\|v_{i z}-w_{i z}\|\right)
\end{aligned}
\end{equation}
\end{small}

where $n$ is the number of Wyckoff positions in the real crystal structure. $v_i$ and $w_i$ are the corresponding atoms in the predicted crystal structure and the real crystal structure. 

\section{Experimental Results}
\label{sec:headings}

\subsection{Generating and screening random crystal structures with a given symmetry}

PyXtal is a Python software package that can generate random structures for a given symmetry and stoichiometry and do structural optimization \cite{2021arXiv210409764Y}.
To predict the crystal structure from multi-objective genetic algorithm, we need a seed structure as starting point to optimize. Here we use PyXtal \cite{fredericks2021pyxtal} to generate candidate template structures.

% \begin{figure}[ht]
%   \centering
%   \includegraphics[width=0.4\linewidth]{figures/PyXtal_process.PNG}
%   \caption{Random structure generation process by PyXtal given the stoichiometry and symmetry}
%   \label{fig:pyxtalstructure}
% \end{figure}

We use the following guidelines to select five suitable random crystal structures from the generated 50 random crystal structures as the template for Wyckoff atomic coordinate optimization:

\begin{itemize}
    \item The number of Wyckoff positions is the least (that is, the multiplicity of each Wyckoff position is as large as possible)
    \item The multiplicity of the Wyckoff positions of each kind of atom is arranged in descending order
    \item The random crystal structures with the top 5 contact map accuracy are selected from the random  structures that meet the above two conditions
\end{itemize}

The Wyckoff position combination of the random crystal structures screened by this method is likely to be consistent with the Wyckoff position combination of the true crystal structure. 

In our algorithm, after successfully screening 5 crystal structures, we will try to search the Wyckoff positions to maximize the match of its contact map and the coordination number of the cation with the target contact map and the coordination number of the cation. 

\begin{table*}[!htb] 
\begin{center}
\caption{Target crystal structures}
\label{table:structure_performance}
\begin{tabular}{|l|l|l|l|l|l|l|l|l|l|}
\hline
\textbf{\makecell{target}} & 
\textbf{\makecell{mp\_id}} & 
\textbf{\makecell{space\\group}} &
\multicolumn{1}{c|}{\textbf{no.of WPs}} &
\textbf{\makecell{No. of atoms\\in unit cell}} &   \multicolumn{1}{c|}{\textbf{\makecell{ coordination number of\\ cations}}}\\ \hline
V\textsubscript{4}S\textsubscript{4}              & mp-1868    & 62               & 2        & 8     & 12         \\ \hline

Zn\textsubscript{4}O\textsubscript{4}              & mp-1093993    & 136               & 2        & 8     & 12         \\ \hline
Pd\textsubscript{4}I\textsubscript{8}              & mp-27747    & 14               & 3        & 12     & 8         \\ \hline

V\textsubscript{6}S\textsubscript{8}              & mp-799    & 176               & 3        & 14     & 12         \\ \hline
Ni\textsubscript{8}O\textsubscript{12}              & mp-1220143    & 9               & 5        & 20     & 12         \\ \hline
As\textsubscript{8}O\textsubscript{12}              & mp-1189365    & 86               & 3        & 20     & 12         \\ \hline
Ce\textsubscript{2}As\textsubscript{2}O\textsubscript{6}              & mp-1078398    & 11               & 4        & 10     & 10         \\ \hline
Cu\textsubscript{4}As\textsubscript{4}S\textsubscript{4}              & mp-5305    & 62               & 3        & 12      & 8   \\ \hline
V\textsubscript{4}Si\textsubscript{4}N\textsubscript{8}              & mp-1246004    & 33               & 4        & 16     & 15         \\ \hline
Li\textsubscript{4}Fe\textsubscript{4}F\textsubscript{16}              & mp-777678    & 14               & 6        & 24     & 16         \\ \hline
\end{tabular}
\end{center}
\end{table*}

\subsection{Prediction results of crystal structure based on multi-objective genetic algorithm}

We selected a set of materials from the Materials Project database as test cases, and the space group numbers are between 9 and 176, as shown in Table \ref{table:structure_performance}. For the NSGA-II algorithm, when optimizing the Wyckoff positions represented as fractional coordinates, we set the range of the variables to [0, 1]. For all the NSGA-II optimizations, the running time ranges from 600 seconds to 3000 seconds, depending on the complexity of the crystal structures. In order to verify the performance of our multi-objective genetic algorithm, we compare the multi-objective genetic algorithm with our previously proposed contact map based CSP algorithm, CMCrystal. The experimental results are shown in Table \ref{table:overall_performance_compare}.

\begin{table*}[!htb] 
\begin{center}
\caption{Prediction performance comparison CMCrystal and Multi-objective genetic algorithm}
\label{table:overall_performance_compare}
\begin{tabular} {|l|l|l|l|l|l|l|l|l|l|l|l|}
% {p{0.15}p{0.05}p{0.05}p{0.05}p{0.05}p{0.05}p{0.05}p{0.05}p{0.05}p{0.05}}
\hline
                    & \multicolumn{4}{c|}{CMCrystal}                                                      & \multicolumn{4}{c|}{Multi-objective GA (CMCrystalMOO)}                                               &\multicolumn{1}{c|}{}                                                              \\ \hline
target           & \multicolumn{1}{c|}{\makecell{contact\\map\\ accuracy}} & 
\multicolumn{1}{c|}{\makecell{coordination\\ error}} &\multicolumn{1}{c|}{RMSD} & \multicolumn{1}{c|}{MAE} & \multicolumn{1}{c|}{\makecell{contact\\map\\accuracy}} &
\multicolumn{1}{c|}{\makecell{coordination\\ error}} &\multicolumn{1}{c|}{RMSD} & \multicolumn{1}{c|}{MAE} &
\multicolumn{1}{c|}{\makecell{Coordination \\error reduction}}  \\ \hline
\makecell{V\textsubscript{4}S\textsubscript{4}} & 1.0                                     &  0                    & 0.243                    & 0.167                     & 1.0                     & 0                    &  0.222           &  0.153                  & 0                  \\ \hline

\makecell{Zn\textsubscript{4}O\textsubscript{4}} & 0.889                                     & 4                     & 0.210                    & 0.163                     & 0.889                     & 2                    &  0.186           &  0.123                  & 2                  \\ \hline
\makecell{Pd\textsubscript{4}I\textsubscript{8}} & 1.0                                     & 4                     &  0.185                   & 0.162                     & 1.0                     & 0                    &  0.136           &  0.110                  & 4                  \\ \hline
\makecell{V\textsubscript{6}S\textsubscript{8}}    & 0.865                                     & 2                     & 0.255                    & 0.148                                          & 0.865                     & 2                    & 0.331                  &  0.229                 & 0                  \\ \hline

\makecell{Ni\textsubscript{8}O\textsubscript{12}}    & 0.933                                     & 5                     & 0.167                    & 0.126                                          & 1.0                     & 1                    & 0.209                  &  0.161                  & 4                  \\ \hline

\makecell{As\textsubscript{8}O\textsubscript{12}}    & 0.8                                     & 3                     & 0.216                    &  0.169                                         & 0.88                     & 0                    & 0.212                  &  0.159                  & 3                  \\ \hline
\makecell{Ce\textsubscript{2}As\textsubscript{2}O\textsubscript{6}} & 0.833                                     & 3                     & 0.298                    & 0.216                                          & 0.9                     & 2                    & 0.227                             &  0.184                  & 1                  \\ \hline
\makecell{Cu\textsubscript{4}As\textsubscript{4}S\textsubscript{4}}       & 0.909                                     & 2                     & 0.221                    & 0.148                                          & 1.0                     & 0                    &  0.208                           & 0.142                   & 2                  \\ \hline
\makecell{V\textsubscript{4}Si\textsubscript{4}N\textsubscript{8}}       & 1.0                                     & 6                     & 0.318                    & 0.275                                          & 1.0                     & 3                    &  0.060                        & 0.050                   & 3                  \\ \hline
\makecell{Li\textsubscript{4}Fe\textsubscript{4}F\textsubscript{16}}    & 0.917                                     & 12                     & 0.285                    & 0.227                                          & 0.957                     & 0                    & 0.268                            & 0.233                  & 12                  \\ \hline
\end{tabular}
\end{center}
\end{table*}

As shown in Table \ref{table:overall_performance_compare}, in the middle\textit{Multi-objective genetic algorithm} columns, we show the crystal structure reconstruction performance of our multi-objective genetic algorithm. For all the test targets, the contact map accuracy range from 0.865 to 1.0 where the lowest accuracy is from V\textsubscript{6}S\textsubscript{8} with the highest space group 176. Accordingly, the coordination number errors of the cation range from 0 to 3,the RMSD errors range from 0.060 to 0.331 and the MAE errors are between 0.050 and 0.233. Compared to CMCrystal's experimental results,we find that for five targets(Ni\textsubscript{8}O\textsubscript{12},As\textsubscript{8}O\textsubscript{12},Ce\textsubscript{2}As\textsubscript{2}O\textsubscript{6}, Cu\textsubscript{4}As\textsubscript{4}S\textsubscript{4} and Li\textsubscript{4}Fe\textsubscript{4}F\textsubscript{16}),the contact map accuracy predicted by the multi-objective genetic algorithm is higher than CMCrystal,others are equal to CMCrystal.This illustrates the excellent performance of our multi-objective genetic algorithm. 
Additionally,we find that for five targets(V\textsubscript{4}S\textsubscript{4},Pd\textsubscript{4}I\textsubscript{8},As\textsubscript{8}O\textsubscript{12}, Cu\textsubscript{4}As\textsubscript{4}S\textsubscript{4} and Li\textsubscript{4}Fe\textsubscript{4}F\textsubscript{16}), we have decreased their coordination number errors of the cation to 0 with 0,4, 3, 2, 12 reduction respectively, which significantly demonstrates the effectiveness of multi-objective genetic algorithm for crystal structure prediction. The all other remaining  targets, the coordination number errors reduction of the cation are between 0 and 4. Moreover, except for the target materials V\textsubscript{6}S\textsubscript{8} and Ni\textsubscript{8}O\textsubscript{12}, the Wyckoff positions coordinates RMSD of other materials predicted by the multi-objective genetic algorithm is less than CMCrystal. In terms of MAE,except for the target materials V\textsubscript{6}S\textsubscript{8}, Ni\textsubscript{8}O\textsubscript{12}  and Li\textsubscript{4}Fe\textsubscript{4}F\textsubscript{16}, the Wyckoff positions coordinates RMSD of other materials predicted by the multi-objective genetic algorithm is less than CMCrystal.

\begin{figure*}[hbt!]
    \centering
    \begin{subfigure}{.24\textwidth}
        \includegraphics[width=\textwidth]{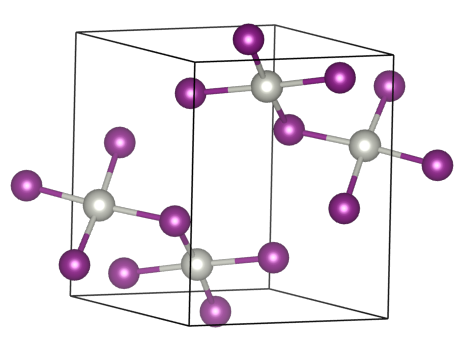}
        \caption{Target structure of Pd\textsubscript{4}I\textsubscript{8}}
        \vspace{3pt}
    \end{subfigure}
    \begin{subfigure}{.24\textwidth}
        \includegraphics[width=\textwidth]{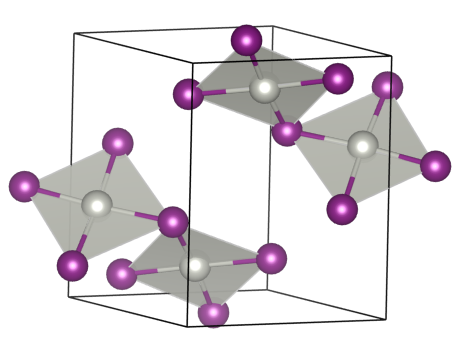}
        \caption{Target structure 
    coordinating polyhedron of Pd\textsubscript{4}I\textsubscript{8}}
        \vspace{3pt}
    \end{subfigure}
    \begin{subfigure}{.24\textwidth}
        \includegraphics[width=\textwidth]{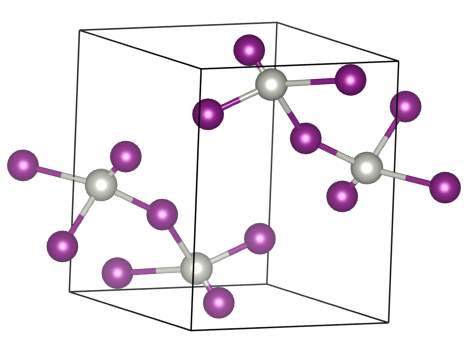}
        \caption{Predicted structure of Pd\textsubscript{4}I\textsubscript{8} with contact map accuracy:100\%, the coordination number error of the cation:0,RMSD: 0.136}
        \vspace{3pt}
    \end{subfigure}
    \begin{subfigure}{.24\textwidth}
        \includegraphics[width=\textwidth]{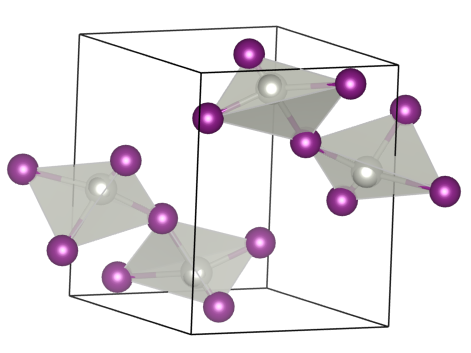}
        \caption{Predicted structure 
    coordinating polyhedron of Pd\textsubscript{4}I\textsubscript{8}}
        \vspace{3pt}
    \end{subfigure}
    \begin{subfigure}{.24\textwidth}
        \includegraphics[width=\textwidth]{figures/V4S4_true.PNG}
        \caption{Target structure of V\textsubscript{4}S\textsubscript{4}}
        \vspace{3pt}
    \end{subfigure}
    \begin{subfigure}{.24\textwidth}
        \includegraphics[width=\textwidth]{figures/V4S4_true_poly.PNG}
        \caption{Target structure 
    coordinating polyhedron of V\textsubscript{4}S\textsubscript{4}}
        \vspace{3pt}
    \end{subfigure}
    \begin{subfigure}{.24\textwidth}
        \includegraphics[width=\textwidth]{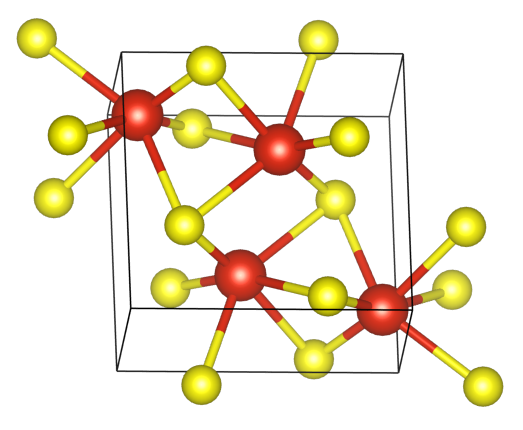}
        \caption{Predicted structure of V\textsubscript{4}S\textsubscript{4} with contact map accuracy:100\%, the coordination number error of the cation:0,RMSD:0.222 }
        \vspace{3pt}
    \end{subfigure}
    \begin{subfigure}{.24\textwidth}
        \includegraphics[width=\textwidth]{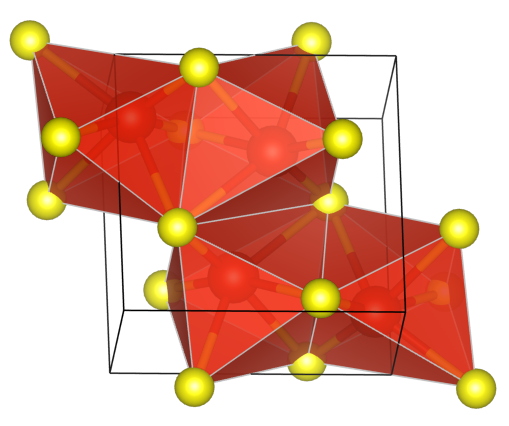}
        \caption{Predicted structure 
    coordinating polyhedron of V\textsubscript{4}S\textsubscript{4}}
        \vspace{3pt}
    \end{subfigure}
    \begin{subfigure}{.24\textwidth}
        \includegraphics[width=\textwidth]{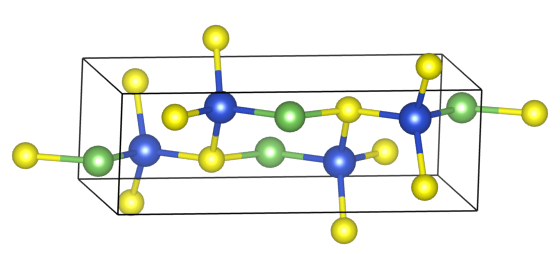}
        \caption{Target structure of Cu\textsubscript{4}As\textsubscript{4}S\textsubscript{4}}
        \vspace{3pt}
    \end{subfigure}
    \begin{subfigure}{.24\textwidth}
        \includegraphics[width=\textwidth]{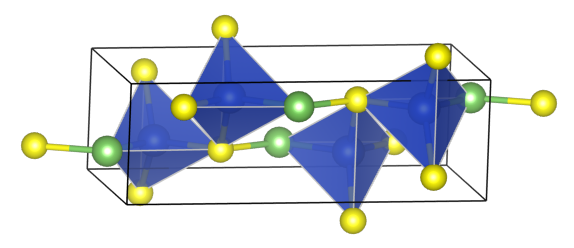}
        \caption{Target structure 
    coordinating polyhedron of Cu\textsubscript{4}As\textsubscript{4}S\textsubscript{4}}
        \vspace{3pt}
    \end{subfigure}
    \begin{subfigure}{.24\textwidth}
        \includegraphics[width=\textwidth]{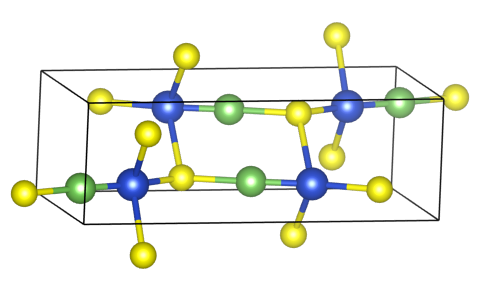}
        \caption{Predicted structure of Cu\textsubscript{4}As\textsubscript{4}S\textsubscript{4} with contact map accuracy:100\%,the coordination number error of the cation:0, RMSD: 0.208}
        \vspace{3pt}
    \end{subfigure}
    \begin{subfigure}{.24\textwidth}
        \includegraphics[width=\textwidth]{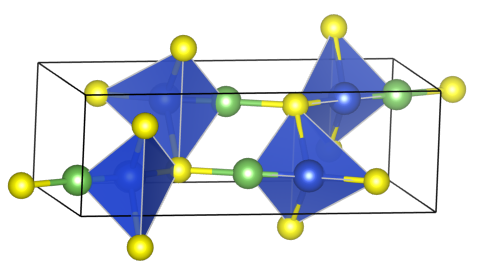}
        \caption{Predicted structure 
    coordinating polyhedron of Cu\textsubscript{4}As\textsubscript{4}S\textsubscript{4}}
        \vspace{3pt}
    \end{subfigure}
    
    \caption{Experimental results of multi-objective genetic algorithm for crystal structure prediction}
    \label{fig:predictedstructures}
\end{figure*}

Figure \ref{fig:predictedstructures} shows three sets of predicted and target crystal structures of Pd\textsubscript{4}I\textsubscript{8}, V\textsubscript{4}S\textsubscript{4}, and Cu\textsubscript{4}As\textsubscript{4}S\textsubscript{4},the contact map accuracy all reache 100\%, the coordination number errors of the cation are all 0 and the predicted structures are very close to the target structures. 

Additionally, we relaxed the predicted structures using DFT based on the  Vienna \textit{ab initio} simulation package (VASP)  \cite{Vasp1,Vasp2,Vasp3,Vasp4}. The plane-wave cutoff energy of 400 eV was considered with the projected augmented wave (PAW) pseudopotentials  \cite{PAW1, PAW2}.  The generalized gradient approximation (GGA)-based exchange-correlation functional was employed by using the Perdew-Burke-Ernzerhof (PBE) method  \cite{GGA1, GGA2}. The energy convergence criterion was 10$^{-5}$ eV, while the force convergence criterion was 10$^{-2}$ eV/{\AA}. The $\Gamma$-centered  Monkhorst-Pack $k$-meshes was used for the Brillouin zone integration.

After the DFT optimization, the formation energies of predicted structures Pd\textsubscript{4}I\textsubscript{8}, V\textsubscript{4}S\textsubscript{4} and Cu\textsubscript{4}As\textsubscript{4}S\textsubscript{4} are -0.018 eV, -0.885 eV, and -0.092 eV respectively.

\section{Discussion}
\label{sec:others}

\begin{comment}
\begin{figure}[H] 
    \begin{subfigure}[t]{0.4\textwidth}
        \includegraphics[width=\textwidth]{figures/}
        \caption{Pd\textsubscript{4}I\textsubscript{8} (after dft optimization) with contact map accuracy:\%, RMSD: }
        \vspace{-3pt}
        \label{fig:K1Ti6Se8_dft}
    \end{subfigure}\hfill
    \begin{subfigure}[t]{0.2\textwidth}
        \includegraphics[width=\textwidth]{figures/}
        \caption{V\textsubscript{4}S\textsubscript{4} (after dft optimization) with contact map accuracy:\%, RMSD: }
        \vspace{-3pt}
        \label{fig:Li2Cr2O4_dft}
    \end{subfigure}
    
    \caption{predicted structures after dft optimization}
    \label{fig:dft}
\end{figure}
\end{comment}

Compared to our previous contact map based crystal structure prediction algorithm CMCrystal,  our multi-objective genetic algorithm CMCrystalMOO takes the contact map, the coordination number of the cations and ages of individuals as the optimization targets, searches for Wyckoff positions, and achieves successful predictions of a set of crystals with high contact map accuracy, low coordination error and low RMSD and MAE errors between the predicted Wyckoff positions and the true Wyckoff positions. However, our evaluations on a large set of target structures show that our algorithm still faces difficulties for In addition to the coordination numbers, other physical patterns such as the polyhedron motif and angle distributions can also be utilized in future work following the physics-informed machine learning paradigm. 

In current experiments, we used the true contact maps, coordination numbers of the cations, space groups and others of the real crystal structures in our crystal structure reconstruction procedure. However, in real situations, the contact map, coordination number of the cation, space group and others are all predicted for a given composition or material formula, which themselves may contain some errors. This may affect the performance of our structure reconstruction algorithm. There are many factors that determine the coordination number of cations, such as the number of period, electric charge, and etc. In many cases, the radius ratio plays an important role. At present, the space group \cite{zhao2020machine,2007Space} and lattice constants \cite{nait2020prediction,li2020mlatticeabc} of crystal structures can be predicted, and the contact map of a given composition can also be predicted by the deep learning  method \cite{hu2021alphacrystal}. With the development of machine learning algorithms, we expect that the various inputs required by our multi-objective genetic algorithm for crystal structure reconstruction can be predicted with high precision, so as to allow our CMCrystalMOO achieves high-quality crystal structure prediction with only a given composition. 

\section{Conclusion}
\label{sec:others}

Improving the scalability of crystal structure prediciton algorithms remains one of the major unsolved problems to make these methods applicable to more complex structures such as ternary materials. We previously proposed CMCrystal, a contact map based crystal structure prediction method, which uses global optimization algorithms such as GA and PSO to search for the Wyckoff positions by maximizing the match between the contact map of the candidate structure and the contact map of the true crystal structure. Our results proved that the geometric constraints in the crystal structure greatly facilitate the reconstruction of the crystal structure. However, our CMCrystal algorithm only uses the connections of the atoms in the unit cell to predict the crystal structure, ignoring the chemical environment outside the unit cell, which may form unreasonable coordination environments. To address this issue and the premature convergence/local optima issue of genetic algorithm, we propose a multi-objective genetic algorithm for contact map base crystal structure reconstruction. We added the optimization goal of the coordination numbers of the cations to CMCrystal algorithm. In order to improve the performance of the optimization process of GA, we take the ages of the individuals in the genetic algorithm as an explicit optimization target. Together, we built a multi-objective crystal structure prediction algorithm based on contact map, coordination number of the cation and ages of the individuals.

We use the contact map accuracy, the coordination number error of the cations, the root mean square distance (RMSD) and the mean absolute error (MAE) between the predicted Wyckoff positions of the crystal structure and those of the real structure to evaluate the quality of the reconstructed crystal structures by our CMCrysalMOO algorithm. Experimental results show that comapred to CMCrystal, our multi-objective crystal structure prediction algorithm can reconstruct the crystal structure with higher quality and can alleviate the problem of premature convergence.

\section{Availability of data}

The data that support the findings of this study are openly available in Materials Project database at \href{http:\\www.materialsproject.org}{\textcolor{blue}{http:\\www.materialsproject.org}}

\section{Contribution}
Conceptualization, J.H.; methodology, J.H., W.Y.; software, W.Y., J.H ; validation, W.Y.,E.S., J.H.;  investigation, J.H., W.Y. E.S.; resources, J.H.; data curation, J.H., and W.Y.; writing--original draft preparation, W.Y. and J.H. ; writing--review and editing, J.H, W.Y., E.S.; visualization, W.Y; supervision, J.H.;  funding acquisition, J.H.

\begin{acknowledgments}
Research reported in this work was supported in part by NSF under grants 1940099 and 1905775. The views, perspective, and content do not necessarily represent the official views of NSF.

\end{acknowledgments}

\bibliography{references}% Produces the bibliography via BibTeX.

\end{document}